\documentclass[final,authoryear,3p,times,twocolumn]{elsarticle}

\usepackage{graphicx}
\usepackage{subfigure}

\usepackage{amsmath}
\usepackage{amssymb}
\usepackage{xcolor}
\usepackage{MnSymbol,wasysym}
\usepackage[T1]{fontenc}

\begin{document}

\begin{frontmatter}



\title{Analog Experiments on Tensile Strength of Dusty and Cometary Matter}


\author[]{Grzegorz Musiolik}
\ead{gregor.musiolik@uni-due.de}
\author[]{Caroline de Beule}
\author[]{Gerhard Wurm}

\address[]{Fakult{\"a}t f{\"u}r Physik, Universit{\"a}t Duisburg-Essen, Lotharstr. 1, 47048 Duisburg, Germany}

\begin{abstract}
The tensile strength of small dusty bodies in the solar system is determined by the interaction between the composing grains. In the transition regime between small and sticky dust ($\rm \mu m$) and non cohesive large grains (mm), particles still stick to each other but are easily separated. In laboratory experiments we find that thermal creep gas flow at low ambient pressure generates an overpressure sufficient to overcome the tensile strength. For the first time it allows a direct measurement of the tensile strength of individual, very small (sub)-mm aggregates which consist of only tens of grains in the (sub)-mm size range. We traced the disintegration of aggregates by optical imaging in ground based as well as microgravity experiments and present first results for basalt, palagonite and vitreous carbon samples with up to a few hundred Pa. These measurements show that low tensile strength can be the result of building loose aggregates with compact (sub)-mm units. This is in favour of a combined cometary formation scenario by aggregation to compact aggreates and gravitational instability of these units.
\end{abstract}

\end{frontmatter}

\section{Introduction}
The solar system swarms with small dusty bodies. Cometary activity close to the sun bears witness to their dusty nature. Small bodies are partly modified by or due to collisions over the last billion of years \citep{krivov2006}.  
However, especially for comets most dust features are likely inherited from the origin within the  solar nebula. Cometary dust was studied in detail by the \textit{Stardust} mission to comet 81P/Wild 2 \citep{Brownlee2006, Hoerz2006, Trigo2008}. More generally, ideas of dusty growth come from experiments or modeling of early phases of planet formation \citep{Blum2008,Johansen2014}. 

Comet flybys of space probes, e.g at Borrelly, Wild 2, and Churyumov-Gerasimenko show that comets are highly porous \citep{Brownlee2006,Davidsson2004a,Davidsson2004b,Patzold2016} and mainly built up of particles with sizes from tens of nanometers to a millimeter \citep{Hoerz2006,Fulle2015a}. With densities of 0.18-0.3 g/cm$^3$, 0.38-0.6 g/cm$^3$, and 0.533 g/cm$^3$, respectively, they are much less dense than pure water ice. High porosities can be the result of quite different internal structures. One extreme might be macropores between otherwise solid, monolithic, large fragments. This resembles a rubble-pile structure. In this case, the building blocks of the pile might be rather solid with an increased strength due to collisional processing and formation of aqueous alteration minerals which fill the pores as e.g. suggested by \citet{Trigo2009}.
The drill on the lander Philae on comet Churyumov-Gerasimenko as part of the \textit{Rosetta} mission, e.g. seemed to have hit solid ground 3 cm below the surface \citep{Spohn2015}. This would be in favour of such ideas. 

Also the dust can provide porosity on differenct scales. On one side there might be very compact dust aggregates, which are then building the larger body again with macro-pore rubble-pile structure. On the other side a homogeneous micro-porosity between micrometer dust grains with low volume filling factor can also provide a global high porosity. An analysis by \citet{Gustafson1996} shows that packing factors of meteors can even be as low as 0.12, which means that the major part of volume might be cavities. 

Some of this structure can be inferred from observations e.g. of the dust production during the active phase or particle entry and ablation into Earth's atmosphere
or IDP analysis \citep{Borovicka1993, Flynn1989, Trigo2003, Fulle2015a, Brownlee1985}. 
Most of the cometary interplanetary dust particles (IDPs) collected in the stratosphere are dense aggregates; typically from 10 $\mu$m to 100 $\mu$m in size and consisting of up to 1 $\mu$m grains \citep{Brownlee1985, Mannel2016}.  
\citet{Fulle2015a} also conclude from measurements of dust ejected from comet Churyumov-Gerasimenko that only a small fraction of the dust is really highly porous.

Related to the morphology, it is the tensile strength which therefore might be a quantity to constrain early planet formation scenarios. For solid monolithic "rocky" material, values for tensile strength are way too large to allow shedding of a dust population. If larger bodies are composed of particles on the order of 1 $\rm \mu m$ in a homogeneous way, their tensile strength is still on the order of 1 kPa \citep{Blum2006}. This assumes that particles only stick together by surface forces and not by chemical bonding, i.e. are not sintered together but that they are packed densely. Such values might be expected from a pure collisional formation of cometesimals with a homogeneous porosity also on large size scales. From meteor observations \citet{Trigo2006,Trigo2007} conclude that cometary particles of sizes from 10$^2$ to 10$^4$ microns entering the atmosphere themselves have tensile strengths of 0.4 to 10 kPa. \citet{Tsuchiyama2009} measured the tensile strength for $\sim$250$\mu$m carbonaceous chondrites to 0.3-30 Mpa.
All this gives clear evidence that on the small scale dust is rather compact and firmly sticks to each other.

Recently discussed are gravitational collapse scenarios for planet (comet) formation where aggregates of mm to cm size form first by collisions in agreement to a large tensile strength. However, these are then concentrated gently and mostly bound together by gravity later \citep{Johansen2014}. In this case, the overall tensile strength of the nucleus' surface is determined by the contacts between these larger granules. The tensile strength measured for such dust granule bodies in analog laboratroy experiments is only at the 1 Pa level \citep{Skorov2012, Blum2014, Brisset2016}. 

These values might be set in the context of processes occuring during  active cometary phases. When comets approach the inner solar system the sublimation of ices ($\rm H_2O$, $\rm CO_2$, CO) leads to a near surface pressure. If this pressure is larger than the tensile strength of the overlaying material, dust is ejected. Sublimation of ice might provide a pressure on the order of a few Pa \citep{Skorov2012}. \citet{Groussin2015} estimate the tensile strength of different parts of comet Churyumov-Gerasimenko to values between a few Pa and 1kPa at maximum. Combined with the measured tensile strength of loosly bound compact aggregates, \citet{Skorov2012,Blum2014,Groussin2015} conclude that these are indications of an early gravitational instability scenario for comets. In any case all these findings indicate the existence of different size scales. Constituents with higher tensile strength are assembled to larger bodies with a lower tensile strength. 

Obviously, the relation between the grain size and the tensile strength is important. Nonetheless, besides the work by \citet{Skorov2012,Blum2014} on large assemblies of dust aggregates there are no dedicated laboratory experiments. Tensile strength has never been explicitly measured before for an individual small aggregate with only few grains within. A question important to judge on formation scenarios would e.g. be: What size distribution of sticky aggregates can build larger assemblies of what tensile strength? This general question is far beyond this paper but we approach this problem here by new laboratory experiments. We developed a technique and used it for the first time to determine the tensile strength of small individual dust aggregates which are only (sub)-mm in size with constituents on the same size scale. 

Our experiments are based on thermal creep gas flow at low pressure. While we use it as technique here to measure tensile strength for small aggregates the interpretation of this work might go beyond. Thermal creep is not restricted to a laboratory environment. Particle disintegration by thermal creep might occur naturally during planet formation and evolution on
the surface of illuminated bodies \citep{kelling2011a, kocifaj2011, debeule2014}. Therefore, this work not only shows laboratory measurements but also allows speculations on the existence of a potentially disastrous process, the knudsen barrier, for weak bodies under certain conditions
\citep{Wurm2007}.

\section{Experimental setup: Ground based}

For the experiments, free moving small aggregates are needed. As we intend to measure low tensile strength, these aggregates are rather fragile by definition. We used two different approaches for our measurements. One setup works in the ground based laboratory. The other setup was used in drop tower experiments. Both are complementary. In this section, we describe the ground based experiments. 

On the ground, free moving aggregates of different sizes can be generated in the following way: At low ambient pressure dust samples are placed on a heater. This leads to thermal creep through the aggregates and an overpressure between the dust and the heater, which levitates dust aggregates \citep{kelling2009}. As this support is adjusting itself to only compensate the weight and is distributing stress over a larger area and partially volume, even low-tensile-strength aggregates can be levitated by this method.
 
Thermal creep is strong enough if the ratio between the mean free path of the gas molecules and the grain or pore size within the aggregate is comparable to or larger than 1. Therefore, low ambient pressure is needed to levitate  the aggregates.\\ 

\begin{figure}[h]
\includegraphics[width=\linewidth]{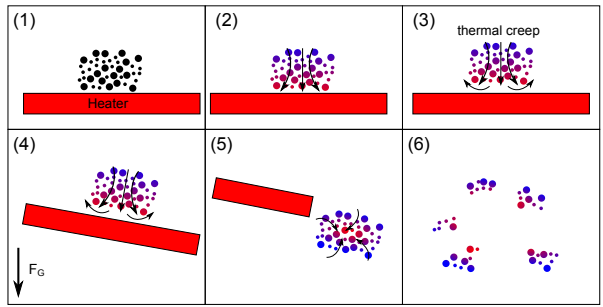}
    \caption{\label{fig.setup1} Principle of the laboratory experiment. A dust sample is placed on a heater within a vacuum chamber (200 Pa) forming aggregates (1). The heater is set to 620 K (2) and based on thermal creep described by \citep{kelling2009}, the aggregates start to levitate over the hot surface (3). The heater is then tilted and the free aggregates slip into free fall (4). Now their outside can cool and a temperature maximum is established inside the aggregate (5). Thermal creep then leads to a gas flow from the cool surroundings to the warm inside. On short timescales, this generates an overpressure and disintegration of the aggregates occurs (6).}
\end{figure}

The main part of the setup thus consists of a heater placed within a vacuum chamber  (Fig.\ref{fig.setup1}). Before the experimental run, the heater is kept at room temperature and for the test experiments basaltic dust with grain sizes smaller than 125 $\mu$m is placed in its center (Fig.\ref{fig.setup1} (1)). The vacuum chamber is evacuated to 200 Pa and the heater is heated to 620 K  (Fig.\ref{fig.setup1} (2)). Eventually, dust aggregates start to levitate over the hot surface (Fig.\ref{fig.setup1} (3)). At this stage, the heater is tilted and the levitating aggregates slip down from the hot surface and are in free fall (Fig.\ref{fig.setup1} (4)). No longer exposed to the hot surface, the aggregates rapidly cool from the outside in due to thermal radiation (Fig.\ref{fig.setup1} (5)). Hence, the maximum temperature is inside the aggregate. In the same way as thermal creep leads to levitation for the aggregates with hot bottom and cool top, gas is now pumped into the aggregate by thermal creep. \\ 

\begin{figure}[h]
	\centering
\includegraphics[width=\linewidth]{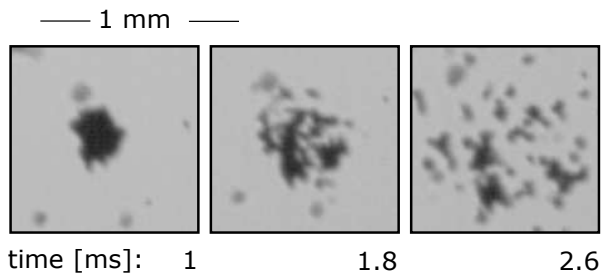}
    \caption{\label{fig.explode}Unprocessed data of a basaltic dust aggregate disintegrating in free fall due to an overpressure induced by thermal creep in the laboratory experiment. }
\end{figure}

On small millisecond timescales, an overpressure builds up inside the aggregate. This effect is also known as Knudsen compressor \citep{knudsen1909}. If the pressure is larger than the tensile strength, the aggregate is explosively disintegrating (Fig.\ref{fig.setup1} (6)). 
The aggregate disintegration is observed by a camera with 25000 frames per second at a resolution of 6 $\rm \mu m$. This temporal and spatial resolution allows to measure the initial acceleration at the start of disintegration. With observed fragment size and known density this directly translates into a tensile strength at the moment the aggregate fragmented. Figure \ref{fig.explode} shows three snapshots of a time sequence of one of such disintegration observed.

\section{Experimental setup: Microgravity}
\begin{figure}[h]
\includegraphics[width=\linewidth]{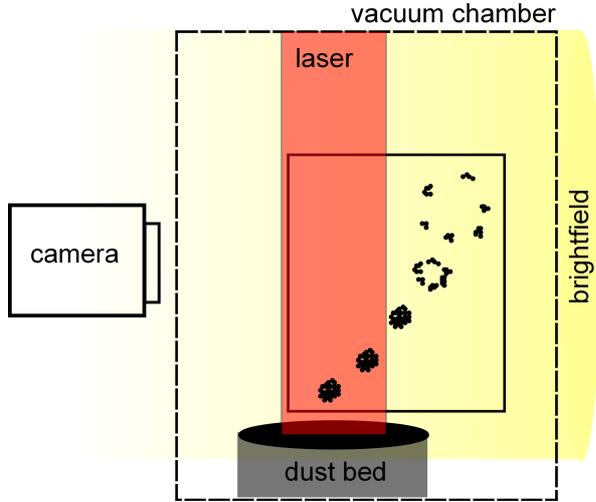}
    \caption{\label{fig.setup} Setup of the microgravity experiment. Suspended dust aggregates are provided by the effect of light induced erosion in microgravity. Once the aggregates left the dust bed and the radiation, they cool from outside-in and an overpressure evolves inside, leading eventually to disintegration.}
\end{figure}

The ground based experiments described above allow an observation at high frame rates and high spatial resolution. Therefore, the accelerated motion upon disintegration can be measured (in 2d) and the values for the tensile strength can directly be determined. Due to the nature of high resolution imaging, only few aggregates which fragment just at the right moment can be observed with this method. No information is gained about a broader aggregate sample, which might disintegrate at locations not observed or which might not fragment at all. To sample a larger number of aggregates it would be beneficial if tensile strengths could be deduced from low resolution imaging of the fragments after the explosion. Then a larger field of view could be observed and statistical
sampling would be possible. We therefore also carried out first drop tower experiments where aggregate fragments can be traced over long timescales of seconds.  \\
A sketch of the experiment is shown in Fig.\ref{fig.setup}.
In this case, suspended aggregates of particle samples were generated by light induced ejections from a particle bed. This ejection mechanism is described, e.g. by \citet{wurm2006, kelling2011a, debeule2013, debeule2014}, and we refer to these papers for details. In the context of this work, it is only important that the released aggregates were part of an illuminated dust bed. An infrared laser beam (955 nm) illuminated a 3.4 cm spot on the dust bed with 7 cm diameter. The light flux was 5.4  and 12.7 kW/m$^2$.  We used basalt and palagonite samples in this work. \\ 
The palagonite was tempered for 1 hour at 900 K to avoid effects of adhering water during the experiments. Furthermore, in a parallel setup a red laser beam (655 nm) with a 5.5 mm spot and a flux of 12.6 kW/m$^2$ was used for a sample of vitreous carbon spheres. \\
\newline
We estimate the temperature of the dust bed to 420 K - 650 K \citep{kocifaj2011}. The sample was in a vacuum chamber at 400 Pa ambient pressure. Once agglomerates leave the surface, they are free to radiate into space and cool from the outside-in, especially if they leave the laser beam. As in the ground based experiments, a radial temperature gradient within the aggregates evolves. \\
The experiments reported here were carried out under microgravity at the drop tower in Bremen. This allows sufficient time for aggregate observations and trajectory reconstruction.

\section{Data analysis}

\subsection{Disintegration at 25000 fps}
\begin{figure}[h]
	\includegraphics[width=\linewidth]{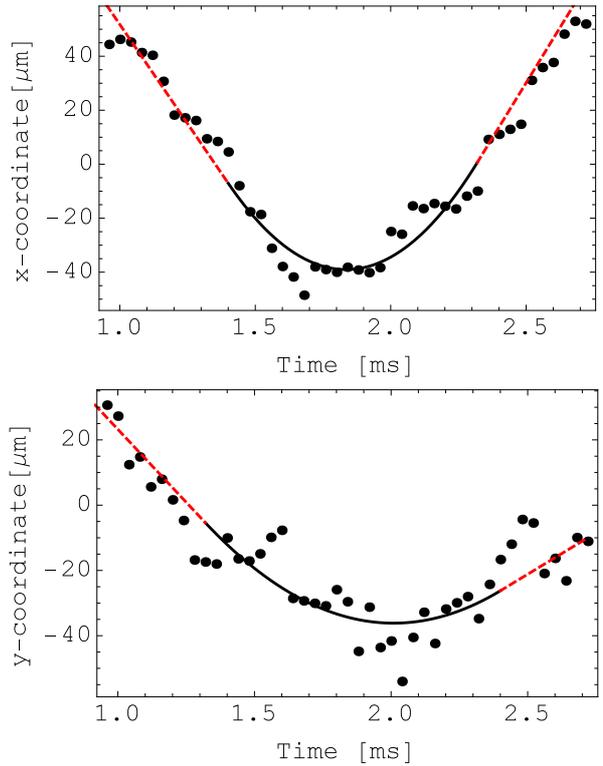}
	\caption{\label{fig.parabola}Example of a fragment position with time in both dimensions observed after a particle disintegration. Gravitational acceleration was subtracted. Overplotted are linear (red, dashed) and parabolic (black, solid) fits.}
\end{figure}

With the given frame rate of 25000 fps in the ground based experiments, the initial acceleration can be resolved which is shown in Fig.\ref{fig.parabola} for one fragment. 
We subtract the gravitational acceleration and rescaled the positions by an arbitrary linear motion to center the trajectory somewhat. The latter is only for visual reasons and is neither showing the trajectory in the center of mass system nor it has an effect on the determination of the acceleration. After removal of the gravity dependence, the trajectory of a fragment can be divided into three distinct parts: a linear motion before disintegration as part of the aggregate, an accelerated part during disintegration and a second linear part after the expansion. Therefore, we approximately fit the resulting curves with two linear parts before and after the explosion with a parabola in between. 
Due to the expansion of the gas within the aggregate the acceleration decreases with time. The approximation of
a parabola with constant acceleration therefore underestimates the initial acceleration 
by a small factor. However, within the accuracy of the data the deviation from a parabola cannot be quantified. We consider the deviation negligible compared to the current uncertainties. We further assume that the velocity is continuous at the connections and set the end points, so that the linear tracks before and after the acceleration have the lowest squared difference to the data. 

The mean force acting on a fragment can be derived from the acceleration data (parabolic part of the track) if the mass of the fragment is known. Taking the cross section of the fragment this also gives a measure of the pressure acting during the explosion which equals the tensile strength. Mass and cross section are estimated from the observed size of the fragment. We approximate the grains by spherical particles of equivalent observed cross section. From this size we estimate the mass by assuming a bulk density of 2.89 g cm$^{-3}$. For 10 fragments analyzed for one fragmentation we get the mean value for the tensile strength
\begin{equation}
\Delta P_\text{acc}= 31.4 \pm 17.4 \; \text{Pa}.
\end{equation}
The observation of such explosions is not trivial and we only consider this as first  measurement here to show the capability of the technique. Also, the data lack a 3d information. Therefore, accelerations are systematically underestimated. 

\subsection{Disintegration at 1000 fps}
\begin{figure}[h]
\includegraphics[width=\linewidth]{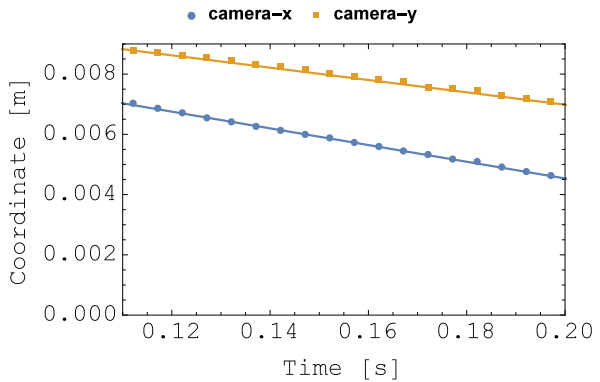}
    \caption{\label{fig.trajectory}Example of a fragment position with time in both dimensions observed after a particle disintegration. x and y refers to camera coordinates. Overplotted are linear fits.}
\end{figure}

\begin{figure}[h]
\includegraphics[width=\linewidth]{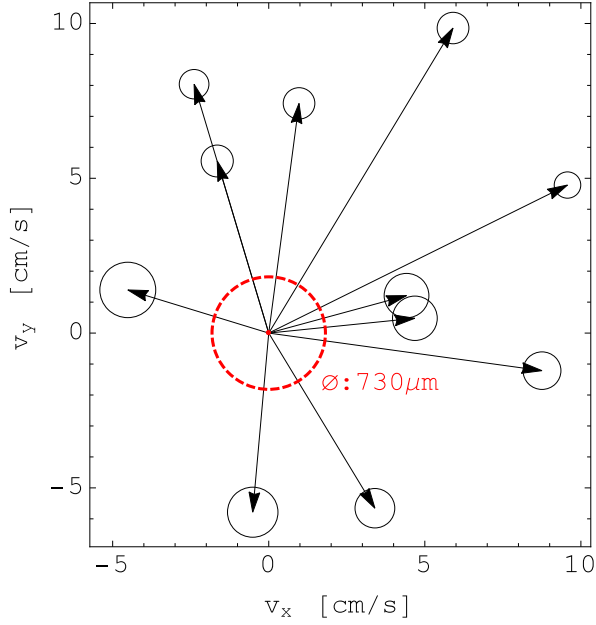}
    \caption{\label{fig.arrows} Velocities (2d projections) and sizes of the fragments in the center of mass system for a disintegration of a basalt aggregate. The sizes of the circles are proportional to the actual sizes of the fragments.}
\end{figure}

\begin{figure}[h]
\includegraphics[width=\linewidth]{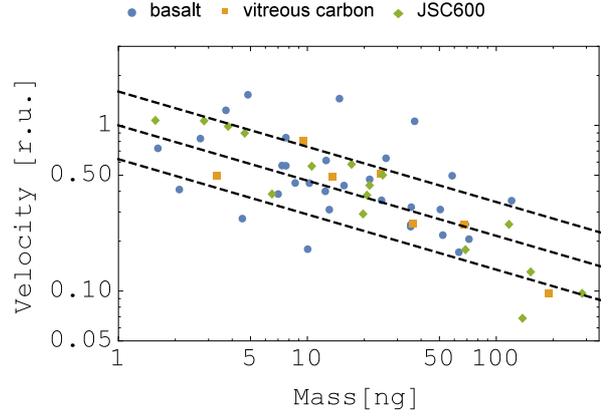}
    \caption{\label{fig.vvonm}2d velocity over mass. The velocities for different disintegrations are scaled to 1 at a mass of 1 ng in
    order to compare aggregates with different absolute velocities. In addition, lines proportional to $m^{-1/3}$ according to a simple explosion model are shown. The upper and lower lines are provided to guide the eye.
    The data cover 7 events (4 for basalt, 2 for palagonite (JSC600) and 1 for carbon glass spheres).}
\end{figure}

\begin{figure}[h]
\includegraphics[width=\linewidth]{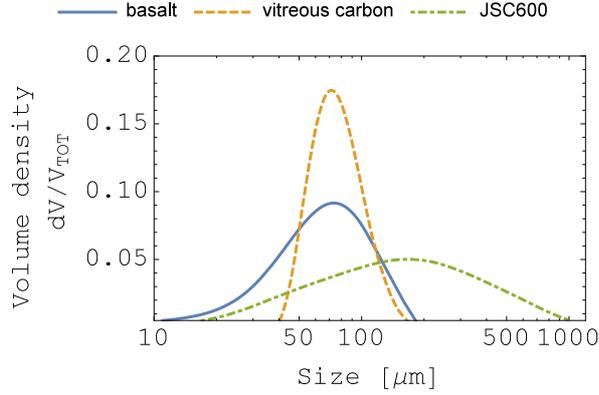}
    \caption{\label{fig.sizedistribution}Volume size distributions in arb. units (probability density but not normalized) of the dust samples studied.}
\end{figure}

During the further analysis we take a look at disintegrating aggregates at 1000 fps from the drop tower experiments. Here, it is no longer possible to resolve the acceleration as shown above. With the help of a simple disintegration model, as detailed below, we nevertheless can estimate the value for the tensile strength. The idea is to determine the kinetic energy of all fragments and compare this to the energy released by the expansion of the gas due to the overpressure.  \\
Fig.\ref{fig.trajectory} shows an example for a recorded fragment trajectory. From the first image, where the particle is observed to 50 images later, the motion can be described by a constant velocity. Note, that there are different scales compared to Fig.\ref{fig.parabola}. All fragments can be described by a linear motion, which allows a straight forward determination of the kinetic energies.
On very long timescales the fragments change their trajectories, as particles couple to the gas and residual flows within the experiment chamber. This is not important here and is not considered further. \\
The motion of the visible center of mass of the aggregate before fragmentation is also determined. The motion of the fragment after the disintegration can then be described in the center of mass system. Fig.\ref{fig.arrows} shows the final data of an explosion reduced to the size and velocity of the fragments in the center of mass system. 
As mentioned above, we define the size of the grains by the radius of a sphere with equivalent observed cross section. To estimate the mass we use a bulk density of 2.89 $\rm g/cm^3$ for basalt, 1.45 $\rm g/cm^3$  for vitreous carbon and 2.5 $\rm g/cm^3$ for the palagonite sample. Adding the mass of all individual grains or smaller aggregates after disintegration often differs from the deduced mass, which can be deduced independently in the same way for the original aggregate. The difference is typically a factor of about 0.5. We attribute the "missing" half of the mass to the porosity of the aggregate with 50\% being a typical value for dust samples consistent with our observation. \\
Fig.\ref{fig.vvonm} shows the velocity of fragments after disintegration of the aggregate over their mass. In total, 7 disintegrations are shown. There is a clear trend that particles with larger mass move slower. The spread has different origins. Particles can move towards or away from the observer and would be misleadingly attributed as slower in a 2d projection. Also, some fragments might pick up rotational energy. Related to this, there is a variation among the fragments in shape which influences the energy that a fragment can get. 
The dotted lines have a slope of $-1/3$ resulting from the simple explosion model as detailed below. A power law of $-1/3$ was fitted to fragments of individual explosions and particle velocities were scaled to a relative velocity of 1 at a mass of 1 ng to compare the different events. 
To complete the data, the size distributions of the dust samples were measured with a Mastersizer 3000 (Malvern Instruments) and are shown in Fig.\ref{fig.sizedistribution}. We note that the basalt sample of the ground based experiment was sieved to below $\rm 125 \mu m$.

\section{Disintegration model}
\begin{figure}[h]
	\centering
    \includegraphics[width=0.99\linewidth]{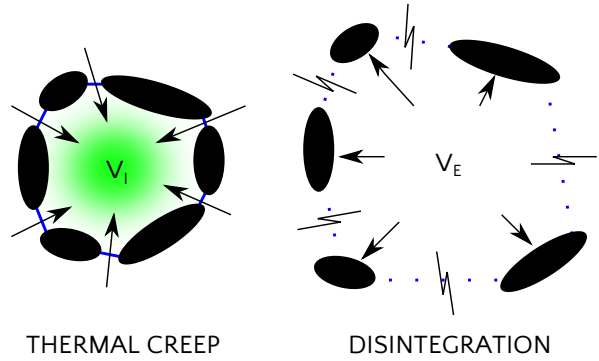}
    \caption{\label{fig.poremodel}Aggregate model with a central pore surrounded by the individual grains and thin capillaries allowing thermal creep to enter the pore.}
\end{figure}

\begin{figure}[h]
\includegraphics[width=\linewidth]{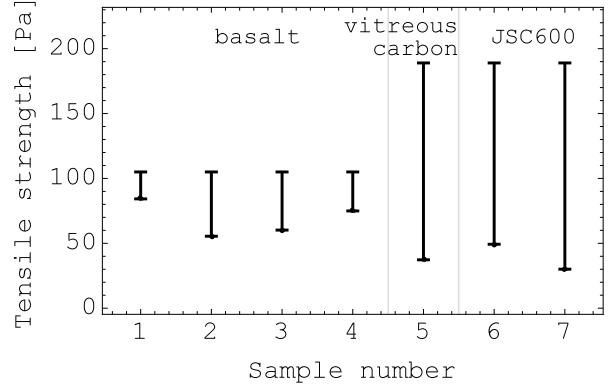}
    \caption{\label{fig.pressures}Tensile strength for different aggregates. The upper limits are the maximum pressure differences from Eq.\eqref{deltap}. The lower limits result from the estimation of the smallest expanding volume $V_E$.}
\end{figure}

In the following, we assume that an overpressure within the aggregate is responsible for the disintegration. It is worth to mention that this occurs on very small spatial scales but 100 $\mu m$ scales for pressure induced tension were also seen in recent work by \citet{debeule2015}. \\
As a model, we approximate an aggregate as a core-mantle structure. The individual grains form a shell surrounding a central pore as seen in Fig.\ref{fig.poremodel}.
If the pressure inside the pore-space increases beyond the tensile strength due to thermal creep, the aggregate (shell) is disrupted and smaller sub-units down to individual grains are accelerated by the pressure as long as the expansion takes place. The maximum pressure difference that can be achieved is given by \citep{knudsen1909}
\begin{equation}
\Delta P_\text{max} = P_{o} \left(  \sqrt{\frac{T_i}{T_o}} -1  \right)
\label{deltap}
\end{equation}
where $P_o$ is the ambient pressure, $T_i$ is the temperature within the pore, and $T_o =$ 300 K is the ambient temperature. $T_i$ depends on the intensity of the corresponding laser and the duration of exposure, which is not exactly known. Assuming values between 1-5 seconds we get $T_i \in [420 \rm K,480 \rm K]$ for the 5.4 kW/m$^2$ laser (used for basalt) and $T_i \in [560 \rm K,650 \rm K]$ for the 12.6 kW/m$^2$ and 12.7 kW/m$^2$ lasers (used for vitreous carbon spheres and JSC-600) referring to \citet{kocifaj2011}. These results lead to overpressures $\Delta P_\text{max} \in [73 \rm Pa,105 \rm Pa]$ for basalt and $\Delta P_\text{max} \in [146 \rm Pa,189 \rm Pa]$ for JSC-600 and the glass-spheres. The real pressure at disintegration can be lower. \\ 
The actual value for the pressure is deduced from the kinetic energy of the fragments which can be measured. Keeping in mind that velocities $v_i$ for fragments with masses $m_i$ are 2d projections only, we get the lower limit of the total kinetic energy in the center of mass system
\begin{equation}
\Delta E_\text{kin}= \sum_i \frac{m_i}{2}v_i^2
\end{equation} 
which is on the order of $0.1$ nJ.\\
The pressure release energy can be calculated by 
\begin{equation}
E_\text{pres} = \int {(P_i - P_o) dV}.
\end{equation}
As it is not possible to resolve the pressure change during the expansion in the images taken with 1000 fps, we assume that the shell is expanding by a distance of spatial resolution adiabatically or with a constant $b = P_i V^\kappa$ where $\kappa$ is the isentropic exponent.
This gives
\begin{equation}
E_\text{pres} = \int_{V_I}^{V_E} {\left(\frac{b}{V^{\kappa}} - P_o\right) dV}
=\frac{b\left(V_E^{1-\kappa}-V_I^{1-\kappa}\right)}{1-\kappa} + P_o (V_I-V_E)
\end{equation}
with the initial volume of the pore $V_I$ and the volume of the pore after the expansion $V_E$. The constant $b$ is given by $b=(P_o + \Delta P) V_I^{\kappa}$. Thus we get
\begin{equation}
\Delta P =  \frac{\left[E_\text{pres}-P_o(V_I-V_E)\right] (1-\kappa)}{V_I^{\kappa}\left(V_E^{1-\kappa}-V_I^{1-\kappa}\right)} -P_o
\label{einp}
\end{equation}
in total. As the fragmentation occurs in a dry air environment, we use $\kappa = 1.4$ \citep{Kouremenos}. Assuming an initial pore size between $[0.3 R_A,0.8 R_A]$ with the radius of the aggregate $R_A$ this allows to estimate $V_I$. In order to calculate $V_E$ we consider radii between $[R_A+ 17 \mu m, R_A + 188 \mu m]$, which is an estimation based on the recorded dataset for 25000 fps ground based experiments. It represents the range of determined expansion volumes, where fragments are accelerated. Finally, Eq.\eqref{einp} allows us to calculate the tensile strength as  $E_\text{pres}= \Delta E_\text{kin}$. Here we assume that the volume work $E_\text{pres}$ is fully transferred into kinetic energy of the fragments. The results are shown in Fig.\ref{fig.pressures}.
If the process was not fully adiabatic, only a fraction of $E_\text{pres}$ could be converted into volume work and thus also into kinetic energy. With this, the tensile strength $\Delta P$ could be rather overestimated by the model due to $\Delta P \sim E_\text{pres}$.

As every fragment is accelerated by the same pressure, the acceleration is proportional to $1/r$ with a fragment size $r$. As all fragments are accelerated over the same time, also the final (measured) velocities are depending on $1/r$ or $1/m^{1/3}$. The mass dependence of the ejecta velocity seen in Fig.\ref{fig.vvonm} is in agreement with this.\\
For the ground based experiments we can apply both methods. As seen above we can directly measure the acceleration and deduce a tensile strength of $\Delta P_\text{acc}= 31.4 \pm 17.4 \; \text{Pa}.$. Using the energy method we get a tensile strength of $\Delta P_\text{E}= 44.7^{+60.3}_{-30.1} \text{Pa}$ for the same aggregate. This is in good agreement and the average values only differ by a factor 1.4. \\

\section{Conclusion}
We measured the tensile strength of small aggregates consisting of basalt, JSC and vitreous carbon on the order of $\sim$ 100Pa for 10 individual grains in the size range of 50 $\rm \mu m$  to 500 $\rm \mu m$ for the first time. We used high as well as low temporal and spatial resolution. We showed that both methods produce similar values. 
Therefore, low resolution imaging with a wider field of view can provide statistical information on tensile strength of a larger aggregate sample in future experiments. This is the first proof of concept that low tensile strength of particle aggregates can be measured.
This technique allows to quantify values in the range expected for cometary matter. The measured tensile strength is on the order of 10 Pa to 100 Pa for three different materials.

While these are first measurements they show that a size scale of 100 $\mu$m is possible as constituent to build the weak surface of comets. These constituents themselves can be more stable sub-units. 
In general, the results are consistent with former work on meteroid fragmentation and strength measurements by the \textit{Stardust} mission of 81P/Wild 2 \citep{Brownlee2006, Trigo2006, Trigo2007, Hoerz2006}. The results fit also the idea of a gravitational instability formation scenario for comets where compact aggregates with high tensile strength grow first and are then assembled to weak larger bodies by gravity as suggested by \citet{Blum2014,Groussin2015,Skorov2012}. It has to be kept in mind that
a minimum aggregate size is needed to concentrate them. E.g. \citet{Bai2010, Carrera2015, Drazkowska2014} show that Stokes numbers (ratio between gas-grain friction time and orbital period) larger than $10^{-2}$ are needed for streaming instabilities though this might 
even be reduced further \citep{Yang2016}. This number depends on the disk model but especially in the comet forming region in the outer solar system this might rather correspond to cm or larger aggregates.
Our results show that the tensile strength is already low for much smaller $100 \rm \mu m$ constituents.
So while different tensile strength for the constituent and the aggregate is in support
of instabilities, there is a mismatch in size scales. In a probably simplified view, the interpretation is that either
gravitational instabilities did not form comets or instabilities were able to concentrate much smaller grains than previously expected.

\section{Acknowledgements}
This project was supported by DLR Space Management with funds provided by the Federal Ministry of Economics Affairs and Energy (BMWi) under grant number DLR 50 WM 1242. G. Musiolik is funded by the DFG. We thank B. Steffentorweihen for help with the lab experiments. We also appreciate the reviews of the two referees.


\begin{thebibliography}{00}

\bibitem[Blum et al.(2006)]{Blum2006} Blum, J., Schr{\"a}pler, 
R., Davidsson, B.~J.~R., \& Trigo-Rodr{\'{\i}}guez, J.~M.\ 2006, ApJ, 652, 1768 

\bibitem[Blum \& Wurm (2008)]{Blum2008} 
Blum, J., \& Wurm, G., 2008, ARAA, 46, 21-56 

\bibitem[Blum et al.(2014)]{Blum2014} 
Blum, J., Gundlach, B., M{\"u}hle, S., \& Trigo-Rodriguez, J.~M., 2014, Icarus, 235, 156 

\bibitem[Bai \& Stone (2010)]{Bai2010} 
Bai, X.-N. \& Stone J.M., 2010, ApJL, 722, L220.

\bibitem[Borovicka (1993)]{Borovicka1993} 
Borovicka, J., 1993, Astronomy \& Astrophysics, 279, 627-645.

\bibitem[Brisset et al.(2016)]{Brisset2016} 
Brisset, J. \& Heisselmann, D. \& Kothe, S. \& Weidling, R. \& 
Blum, J., 2016, Astronomy \& Astrophysics 593, A3

\bibitem[Brownlee et al.(1985)]{Brownlee1985} 
Brownlee, D., 1985, Ann. Rev. Earth Planet. Sci. 13, 147-173

\bibitem[Brownlee et al.(2006)]{Brownlee2006} 
Brownlee, D.\& Tsou, P.\& Aléon, J.\& Alexander, C. M. D.\& Araki, T.\& Bajt, S., ... \& Borg, J., 2006, Science, 314(5806), 1711-1716.

\bibitem[Carrera et. al (2015)]{Carrera2015} 
Carrera, D. \& Johansen, A. \& Davies, M. B., 2015, Astronomy \& Astrophysics, 579, A43

\bibitem[Davidsson \& Guti{\'e}rrez(2004)]{Davidsson2004a} 
Davidsson, B.~J.~R., \& Guti{\'e}rrez, P.~J.\ 2004, Icarus, 168, 392 

\bibitem[Davidsson \& Gutierrez(2004)]{Davidsson2004b} 
Davidsson, B.~J.~R., \& Gutierrez, P.~J.\ 2004, Bulletin of the American Astronomical Society, 36, 1118 

\bibitem[de Beule et al.(2013)]{debeule2013} 
de Beule, C., Kelling, T., Wurm, G., Teiser, J., \& Jankowski, T., 2013, ApJ, 763, 11 

\bibitem[de Beule et al. (2014)]{debeule2014}
de Beule, C., Wurm, G. and Kelling, T., K{\"{u}}ppers, M., Jankowski, T. \& Teiser, J., 2014, Nature, 10

\bibitem[{de Beule} et al., 2015]{debeule2015}
de Beule, C., Wurm, G., Kelling, T., Koester, M., Kocifaj, M., 2015, Icarus, 260, 23

\bibitem[Dominik \& Tielens, 1997]{dominik1997}
Dominik, C., Tielens, A.G.G.M., 1997, ApJ, 480:647-673

\bibitem[Dr\k{a}\.{z}kowska \& Dullemond, 2014]{Drazkowska2014}
Dr\k{a}\.{z}kowska, J. \& Dullemond, C. P., 2014, Astronomy \& Astrophysics, 572, A78.

\bibitem[{Flynn}, 1989]{Flynn1989}
Flynn, G. J., 1989, Icarus, 77(2), 287-310.

\bibitem[{Fulle} et~al., 2015]{Fulle2015a}
Fulle, M., Della Corte, V., Rotundi, A., et al., 2015, ApJ, 802, L12

\bibitem[{Groussin} et~al., 2015]{Groussin2015}
Groussin, O., Jorda, L., Auger, A. T., Kührt, E., Gaskell, R., Capanna, C., ... \& Knollenberg, J., 2015, Astronomy \& Astrophysics, 583, A32.

\bibitem[{Gustafson} \& {Adolfsson}, 1996]{Gustafson1996}
Gustafson, B.\r{A}. S., and L. G. Adolfsson \ 1996, The Cosmic Dust Connection. Springer Netherlands, 349-355.

\bibitem[{H{\"o}rz} et~al., 2006]{Hoerz2006}
H{\"o}rz, F., Bastien, R., Borg, J., Bradley, J. P., Bridges, J. C., Brownlee, D. E., ... \& Djouadi, Z., 2006, Science, 314(5806), 1716-1719.

\bibitem[Johansen et al.(2014)]{Johansen2014} 
Johansen, A., Blum, J., Tanaka, H., et al.\ 2014, Protostars and Planets VI, 547 

\bibitem[Kelling \& Wurm, 2009]{kelling2009}
Kelling, T. \& Wurm, G. 2009, Phyis. Rev. Lett., 103, 21

\bibitem[{Kelling} et~al., 2011]{kelling2011a}
{Kelling}, T., {Wurm}, G., {Kocifaj}, M., {Kla{\v c}ka}, J., \& {Reiss}, D., 2011, Icarus, 212, 935

\bibitem[Knudsen(1909)]{knudsen1909} 
Knudsen, M.\ 1909, Annalen der Physik, 336, 205 

\bibitem[{Kocifaj} et~al., 2011]{kocifaj2011}
{Kocifaj}, M., {Kla{\v c}ka}, J., {Kelling}, T., \& {Wurm}, G. 2011, Icarus, 212, 935

\bibitem[{Kouremenos} et~al., 1986]{Kouremenos}
Kouremenos, D.A., Acta Mechanica, 1987, 1-4, 81-99

\bibitem[Krivov et al.(2006)]{krivov2006} 
Krivov, A.~V., L{\"o}hne, T., \& Srem{\v c}evi{\'c}, M.\ 2006, Astronomy \& Astrophysics, 455, 509 

\bibitem[Mannel et al.(2016)]{Mannel2016} 
Mannel, T. \& Bentley, M. S. \& Schmied, R. \& Jeszenszky, H. \& Levasseur-Regourd, A. C. \& Romstedt, J. \& Torkar, K. 2016, Monthly Notices of the Royal Astronomical Society, 462, S304

\bibitem[P{\"a}tzold et al.(2016), 2016]{Patzold2016}
P{\"a}tzold, M. \& Andert, T. \&, Hahn, M. \&, Asmar, S.W. \& Barriot, J.-P.\&	Bird, M. K. \&	 H{\"a}usler, B.\& Peter, K. \& Tellmann, S. \&	Gr{\"u}n, E. \& Weissman, P. R.\& Sierks, H. \& Jorda, L. \& Gaskell, R. \& Preusker F. \& Scholten, F., Nature, 530, 63-65

\bibitem[Skorov \& Blum, 2012]{Skorov2012}
Skorov, Y. V. \& Blum, J. 2012, Icarus, 221, 1

\bibitem[Spohn et al.(2015)]{Spohn2015}
Spohn, T., Knollenberg, J., Ball, A.~J., et al.\ 2015, Science, 349, 020464 

\bibitem[Trigo-Rodr{\'{\i}}guez et al.(2003)]{Trigo2003} 
Trigo-Rodr{\'{\i}}guez, J. M., Llorca, J., Borovicka, J. \& Fabregat, J. 2003, Meteoritics \& Planetary Science, 38(8), 1283-1294.

\bibitem[Trigo-Rodr{\'{\i}}guez \& Llorca (2006)]{Trigo2006} 
Trigo-Rodr{\'{\i}}guez, J.M., \& Llorca, J.\ 2006, Monthly Notices of the Royal Astronomical Society, 372, 655 

\bibitem[Trigo-Rodr{\'{\i}}guez \& Llorca (2007)]{Trigo2007} 
Trigo-Rodr{\'{\i}}guez, J.M., \& Llorca, J.\ 2007, Monthly Notices of the Royal Astronomical Society, 375, 415 

\bibitem[Trigo-Rodr{\'{\i}}guez et al.(2008)]{Trigo2008} 
Trigo-Rodr{\'{\i}}guez, J. M.\& Dom{\'{\i}}nguez, G.\& Burchell, M. J.\& H{\"o}rz, F. \& Llorca, J.\ 2008, Meteoritics \& Planetary Science, 43: 75–86

\bibitem[Trigo-Rodr{\'{\i}}guez \& Blum (2009)]{Trigo2009} 
Trigo-Rodr{\'{\i}}guez, J. M. \& Blum, J. \ 2009, Planetary and Space Science 57.2: 243-249.

\bibitem[{Tsuchiyama} et al. (2009)]{Tsuchiyama2009} 
{Tsuchiyama}, A. \& {Mashio}, E. \& {Imai}, Y. \& {Noguchi}, T. \& 
{Miura}, Y. \& {Yano}, H. \& {Nakamura}, T., 2009, Meteoritics and Planetary Science Supplement, 72, 5189 

\bibitem[{Wurm} \& {Krauss}, 2006]{wurm2006}
{Wurm}, G. ,\& {Krauss}, O. 2006, Phys. Rev. Lett., 96, 134301

\bibitem[{Wurm}, 2007]{Wurm2007}
{Wurm}, G. ,\ 2007, Monthly Notices of the Royal Astronomical Society,  380, 683-690

\bibitem[Yang et al., 2016]{Yang2016}
{Yang}, C.-C. ,\& {Johansen}, A. ,\& {Carrera}, D. 2016, arXiv preprint arXiv:1611.07014 (2016).





 \end{thebibliography}



\end{document}